\documentclass[ams,aps,10pt]{revtex4}

\usepackage{amsmath}
\usepackage{amssymb}

\begin{document}

\newcommand{\TeXButton}[2]{#2}

\title{Orthogonality relations for triple modes at dielectric boundary
surfaces}

\author{Hanno Hammer}%
\email{H.Hammer@umist.ac.uk}
\affiliation{Department of Chemical Physics,\\
             The Weizmann Institute of Science,\\
             76100 Rehovot, Israel}

\date{30. Nov. 2001}

\begin{abstract}
We work out the orthogonality relations for the set of
Carniglia-Mandel triple modes which provide a set of normal modes for
the source-free electromagnetic field in a background consisting of a
passive dielectric half-space and the vacuum, respectively. Due to the
inherent computational complexity of the problem, an efficient
strategy to accomplish this task is desirable, which is presented in
the paper. Furthermore, we provide all main steps for the various
proofs pertaining to different combinations of triple modes in the
orthogonality integral.
\end{abstract}


\maketitle


\section{Introduction}

The electromagnetic field propagating between boundary surfaces of
conducting or dielectric media differs from the free electromagnetic
field in vacuum in that the boundary conditions imposed by the
material media on the field alter the structure of the normal modes of
the field as well as the state density of electromagnetic field
modes. The alteration of these quantities becomes manifest when
transition rates and energy level shifts of atomic matter exposed to
the field environment are studied, as these quantities directly enter
the expressions of the relevant transition matrix elements.

The complex of problems relating to these ideas is now generally
called Cavity Quantum Electrodynamics (CQED). One of the principal
ideas underlying CQED appeared first in the abstract of paper
\cite{Purcell}. CQED as a subject has been initiated in
\cite{JaynesCummings1963} with the study of atoms coupled to a single
quantized mode of the radiation field in a cavity, followed by
articles \cite {Stehle1970} and \cite{Barton1970} working on
modifications of spontaneous decay rates and radiative level shifts of
optical transitions in a cavity.  Resonators, waveguides, dielectrics,
and conducting interfaces modifying atomic radiative properties such
as decay times, shifts in energy levels, and spectral line shapes have
been reviewed in \cite{Hinds1}. A review of the subject of CQED in the
presence of metallic boundaries including exhaustive references to
original papers was given in \cite{Berman1994}.

The mode structure as well as the problem of atomic excitation and
photoemission by evanescent waves at a dielectric half-space have been
derived at first in \cite{CarnigliaMandel1,CarnigliaMandel2}. There it
was found that an appropriate set of normal modes was given by
so-called triple modes, which are composed of three plane-wave
components (incoming, reflected, transmitted). The transmitted
component of a triple mode can be a travelling wave or an evanescent
wave, with exponentially decreasing amplitude behind the boundary
plane. A variant of these modes has been used in \cite{JanZak1994},
where the radiation properties of a quantum harmonic oscillator near a
planar dielectric half-space were examined. In \cite {InoueHori2001},
special linear combinations of time-reversed and spatially rotated
Carniglia-Mandel triple modes have been employed to determine
spontaneous emission rates and final-state mode densities. A further
generalization has been presented in \cite
{UrbachRikken1,ZakowiczBledowski1995}, where spontaneous emission in
the presence of a dielectric slab bounded by two dielectric
half-spaces was studied. General quantum optics with the main focus on
quantum fluctuations in dielectric media was discussed in
\cite{GlauberLewenstein1991}.

One of the central problems in these developments is the determination
of an appropriate set of normal modes of the free electromagnetic
field subject to the given boundary conditions. These normal modes
obey the source-free Maxwell equations in a dielectric and satisfy the
boundary conditions describing either conducting or dielectric 
surfaces. Having solved for a set of modes it is crucial to check
their orthogonality, as this property determines whether the free
field Hamiltonian can be diagonalized in these modes.

The question of orthogonality of normal modes is simple only in the
case of no boundaries, i.e. the field in vacuum. As soon as boundaries
are imposed, checking for orthogonality can be a tedious
procedure. For this reason, actual computations proving the
orthogonality of modes currently used in the literature are not widely
available. For example, even in the seemingly simple problem of a
dielectric half-space the question of orthogonality of the triple
modes becomes highly non-trivial. In particular, orthogonality of a
pair of triple modes containing a traveling and an evanescent
component in the same half-space, respectively, is quite
counter-intuitive. A brief proof of orthogonality in the case of a
dielectric half-space has first been given in the appendix of
\cite{CarnigliaMandel1}. We feel, however, that there is a demand for
a more detailed examination of this problem.

Our work presented here is intended to fill this gap. We expand on the
results presented in the appendix of \cite{CarnigliaMandel1}, and give
a proof of orthogonality for all different cases of triple modes as
they appear in the dielectric half-space problem.  Each proof attempts
to highlight the most important intermediate steps as well as the
mathematical tools necessary to derive them.  We do not enter the
issue of quantization of these modes, but will end by showing that the
free field Hamiltonian can be diagonalized in these modes and hence
can be written as a sum of independent harmonic oscillators pertaining
to these triple modes, within a classical context. The completeness of
the set of triple modes in the space of transverse vector-valued
functions has been proven in \cite{BB1972}. This, together with our
result, shows that the triple modes are actually 'good' modes for a
canonical quantization of the electromagnetic field in a dielectric
half-space.

In this approach, the triple modes appear as traveling normal modes of
the free electromagnetic field. The term 'free' deserves some
explanation here: Strictly speaking, propagation of an electromagnetic
plane wave through a material medium is the result of the interaction
of the fundamental fields ${\bf E}$ and ${\bf B}$ with a macroscopic
number of microscopic sources constituting the bulk matter. In
principle, these sources must be incorporated into the dynamics of the
total system by suitable interaction terms. However, if the frequency
band width of the plane waves under consideration is far off the
resonance frequencies of the bulk matter, any light-matter interaction
will be only transient in nature, and will certainly not affect the
macroscopic state of the bulk matter. In this case, we can treat the
bulk matter as a passive medium, whose macroscopic properties with
respect to the interaction with radiation can be summarized by
introducing a (space- and time-dependent) refractive index. In the
following we will use this approach.

\section{Triple modes}

The Carniglia-Mandel modes are assumed to propagate in a background
which consists of a non-magnetic material medium with refractive index
$n=\sqrt{\frac{\epsilon \mu _0}{\epsilon _0\mu _0}}$ in the left
half-space $z<0$, and the vacuum with refractive index $n=1$ in the
right half-space $z>0$, where $\epsilon $, $\epsilon _0$ are the
permittivities of the medium and the vacuum, respectively, and $\mu
_0$ is the uniform magnetic permeability which takes the vacuum value
everywhere, since the medium is non-magnetic. Furthermore, $\epsilon $
is assumed to be time-independent.

Before we state the form of the triple modes, we introduce our
notation conventions. Let ${\bf e}_3$ be the unit normal vector onto
the boundary plane (i.e. the plane extends in the $xy$-directions),
then any vector ${\bf V}$ on the boundary can be decomposed into
components parallel and perpendicular to the boundary plane ${\bf
V}={\bf V} _{\parallel }+{\bf V}_{\perp }$. In particular, the wave
vector ${\bf k}$ of a homogeneous plane wave incident on the boundary
plane can be decomposed as ${\bf k=}K{\bf e}_3{\bf +k}_{\parallel }$,
where ${\bf k}\bullet {\bf e} _3\equiv K$. In the following, this
notation will be used extensively.

The triple modes are obtained by making a separation ansatz for the
electric/magnetic fields, factoring out the time dependence in the form of $%
e^{-i\omega t}$. A typical electric field will then take the form ${\bf E}%
\left( {\bf k},{\bf x}\right) e^{-i\omega t}$, where the spatial part ${\bf E%
}\left( {\bf k},{\bf x}\right) $ depends on the wave vector ${\bf k}$ and
satisfies an appropriate Helmholtz equation.

The triple modes can be distinguished by their polarization with respect to
the incident plane, which is spanned by the incoming wave vector ${\bf k}_i$
and the normal vector ${\bf e}_3$. If the incoming electric field ${\bf E}_i$
is perpendicular to the incident plane, the mode is called
transverse-electric (TE); this will be indicated by polarization index $1$.
If the incoming electric field lies in the plane of incidence, the mode is
called transverse-magnetic (TM), indicated by polarization index $2$. Each
triple mode is composed of an incoming, a reflected and a transmitted plane
wave component. We shall label all quantities pertaining to these components
by subscripts $i$, $r$, $t$, respectively. Furthermore, fields will carry a
subscript $L$ or $R$ depending on whether the corresponding incoming wave
comes from the left, $K_i>0$, or from the right, $K_i\le 0$.

To simplify the subsequent computations we first choose normalization
conventions different from those in \cite{CarnigliaMandel1}: we assign
unit amplitude to incoming electric plane waves in the TE case, and
unit amplitude to incoming magnetic plane waves in the TM case. Only
at the end of our work shall we discuss a different normalization 
appropriate for the diagonalization of the field Hamiltonian.

The TE modes incoming from the left are labelled by $L$, the incoming
wave vector ${\bf k}_i$, and the polarization index $1$, and are given by
their electric field components 
\begin{equation}
\label{te1}{\bf E}_L\left( {\bf k}_i,1,{\bf x}\right) =\left\{ 
\begin{array}{ccc}
{\bf e}\left[ e^{i{\bf k}_i\bullet {\bf x}}+a_re^{i{\bf k}_r\bullet {\bf x}%
}\right] & , & z<0 \\ 
{\bf e}a_te^{i{\bf k}_t\bullet {\bf x}} & , & z\ge 0 
\end{array}
\right. \quad ,
\end{equation}
where the real unit polarization vector ${\bf e}$ is chosen in the direction
of ${\bf k}_i\times {\bf e}_3$, and the incoming wave is assumed to have
unit amplitude. ${\bf k}_r$ is the wave vector of the reflected wave, with $%
K_r=-K_i$, and ${\bf k}_t$ is the wave vector of the transmitted wave; its
component normal to the boundary $K_t$ can be real or imaginary, depending
on the angle of the incoming wave; in the latter case, the triple mode is
called \underline{evanescent}. The three components of the mode must (together
with their time dependence $e^{-i\omega t}$) satisfy the Maxwell equations
on the boundary \cite{Jackson} as well as a phase matching condition. This
determines the reflexion / transmission coefficients $a_r$, $a_t$ to be 
\begin{equation}
\label{reftrans1}
a_r=\frac{K_i-K_t}{K_i+K_t}\quad ,\quad a_t=\frac{2K_t}{ K_i+K_t}
\quad.
\end{equation}
For modes incoming from the left, frequency $\omega $ and wave vectors
${\bf k}$ are related by
\begin{equation}
\label{reftrans2}{\bf k}_{\Vert }^2+K_i^2={\bf k}_{\Vert }^2+K_r^2=\mu
_0\epsilon \omega ^2\quad ,\quad {\bf k}_{\Vert }^2+K_t^2=\mu _0\epsilon
_0\omega ^2\quad . 
\end{equation}
We note that $K_i$ and $K_r$ are always real, but the fact that
$\epsilon > \epsilon _{0}$ allows for $K_{t}$ to be real or imaginary;
in the latter case, the transmitted wave is evanescent.  The
associated TE \underline{magnetic}\ field is uniquely determined by
Maxwell's equations, and is given by
\begin{equation}
\label{te2}{\bf B}_L\left( {\bf k}_i,1,{\bf x}\right) =\left\{ 
\begin{array}{ccc}
\frac{{\bf k}_i\times {\bf e}}\omega e^{i{\bf k}_i{\bf \bullet x}}+a_r\frac{%
{\bf k}_r\times {\bf e}}\omega e^{i{\bf k}_r{\bf \bullet x}} & , & z<0 \\ 
a_t\frac{{\bf k}_t\times {\bf e}}\omega e^{i{\bf k}_t{\bf \bullet x}} & , & 
z\ge 0 
\end{array}
\right. \quad . 
\end{equation}

Similarly, the TE modes normalized upon unit electric field amplitude
incoming from the right are labeled by $R$, the incoming wave vector
${\bf k}_i$ with $K_i\le 0$, and the polarization index $1$, and are
given by
\begin{equation}
\label{te1r}{\bf E}_R\left( {\bf k}_i,1,{\bf x}\right) =\left\{ 
\begin{array}{ccc}
{\bf e}a_te^{i{\bf k}_t\bullet {\bf x}} & , & z<0 \\ 
{\bf e}\left[ e^{i{\bf k}_i\bullet {\bf x}}+a_re^{i{\bf k}_r\bullet {\bf x}%
}\right] & , & z\ge 0 
\end{array}
\right. \quad ,
\end{equation}
where ${\bf e}$ has the same direction as ${\bf k}_i\times {\bf
e}_3$. The reflexion / transmission coefficients $a_r$, $a_t$ have the
same form as in (\ref {reftrans1}), but the relation between frequency
$\omega $ and wave vectors is now
\begin{equation}
\label{reftrans2r}{\bf k}_{\Vert }^2+K_i^2={\bf k}_{\Vert }^2+K_r^2=\mu
_0\epsilon _0\omega ^2\quad ,\quad {\bf k}_{\Vert }^2+K_t^2=\mu _0\epsilon
\omega ^2\quad . 
\end{equation}
The associated TE \underline{magnetic}\ field is 
\begin{equation}
\label{te2r}{\bf B}_R\left( {\bf k}_i,1,{\bf x}\right) =\left\{ 
\begin{array}{ccc}
a_t\frac{{\bf k}_t\times {\bf e}}\omega e^{i{\bf k}_t{\bf \bullet x}} & , & 
z<0 \\ 
\frac{{\bf k}_i\times {\bf e}}\omega e^{i{\bf k}_i{\bf \bullet x}}+a_r\frac{%
{\bf k}_r\times {\bf e}}\omega e^{i{\bf k}_r{\bf \bullet x}} & , & z\ge 0 
\end{array}
\right. \quad . 
\end{equation}

Now we turn to the TM modes: The \underline{magnetic}\ field of the 
TM modes incoming from the left is 
\begin{equation}
\label{tm1}{\bf B}_L\left( {\bf k}_i,2,{\bf x}\right) =\left\{ 
\begin{array}{ccc}
{\bf e}\left[ e^{i{\bf k}_i\bullet {\bf x}}+b_re^{i{\bf k}_r\bullet {\bf x}%
}\right] & , & z<0 \\ 
{\bf e}b_te^{i{\bf k}_t\bullet {\bf x}} & , & z\ge 0 
\end{array}
\right. \quad , 
\end{equation}
where ${\bf e}$ is a real unit magnetic polarization vector chosen in
the direction of ${\bf k}_i\times {\bf e}_3$. The associated
\underline{electric} field follows from Maxwell's equations, and is
given by
\begin{equation}
\label{tm2}{\bf E}_L\left( {\bf k}_i,2,{\bf x}\right) =\left\{ 
\begin{array}{ccc}
-\frac{{\bf k}_i\times {\bf e}}{\omega \mu _0\epsilon }e^{i{\bf k}_i{\bf %
\bullet x}}-\frac{{\bf k}_r\times {\bf e}}{\omega \mu _0\epsilon }e^{i{\bf k}%
_r{\bf \bullet x}} & , & z<0 \\ 
-\frac{{\bf k}_t\times {\bf e}}{\omega \mu _0\epsilon _0}e^{i{\bf k}_t{\bf %
\bullet x}} & , & z\ge 0 
\end{array}
\right. \quad . 
\end{equation}
The relation between frequency and wave vectors is given in eq
(\ref{reftrans2} ). For TM waves incoming from the right we have
\begin{equation}
\label{tm1r}{\bf B}_R\left( {\bf k}_i,2,{\bf x}\right) =\left\{ 
\begin{array}{ccc}
{\bf e}b_te^{i{\bf k}_t\bullet {\bf x}} & , & z<0 \\ 
{\bf e}\left[ e^{i{\bf k}_i\bullet {\bf x}}+b_re^{i{\bf k}_r\bullet {\bf x}%
}\right] & , & z\ge 0 
\end{array}
\right. \quad , 
\end{equation}
with associated electric field 
\begin{equation}
\label{tm2r}{\bf E}_R\left( {\bf k}_i,2,{\bf x}\right) =\left\{ 
\begin{array}{ccc}
-\frac{{\bf k}_t\times {\bf e}}{\omega \mu _0\epsilon }e^{i{\bf k}_t{\bf %
\bullet x}} & , & z<0 \\ 
-\frac{{\bf k}_i\times {\bf e}}{\omega \mu _0\epsilon _0}e^{i{\bf k}_i{\bf %
\bullet x}}-\frac{{\bf k}_r\times {\bf e}}{\omega \mu _0\epsilon _0}e^{i{\bf %
k}_r{\bf \bullet x}} & , & z\ge 0 
\end{array}
\right. \quad . 
\end{equation}
The relation between wave vectors and frequency is given in eq
(\ref{reftrans2r}).

The reflexion / transmission coefficients $b_r$ and $b_t$ now take a
slightly more complicated form in the TM case. In order to avoid the
need to explicitly distinguish between left- and right-propagating
modes, we introduce the following notation conventions: Let
$\mathbb{H}_i$ be the half-space in which the \underline{incoming}
component of the triple mode under consideration propagates, and let
$\varepsilon _i$ and $n_i$ be the permittivity and refractive index of
this half-space; similarly, denote $\mathbb{H}_t$ for the half-space
which contains the transmitted component, with associated permittivity
$\varepsilon _t$ and refractive index $n_t$.
 
Thus, for waves incoming from the left we have $\mathbb{H}_i=\{z<0\}$
and $\mathbb{H}_t=\{z\ge 0\}$, as well as $\epsilon _i=\epsilon $,
$\epsilon _t=\epsilon _0$, and $n_i=n_0$, $n_t=1$; for waves incoming
from the right we must interchange $i$ and $t$ in these formulas.
With this convention, the reflexion / transmission coefficients for
the TM waves can be written in a uniform way as
\begin{equation}
\label{mo27}b_r=\frac{K_i/\epsilon _i-K_t/\epsilon _t}{K_i/\epsilon
_i+K_t/\epsilon _t}\quad ,\quad b_t=2\frac{K_i/\epsilon _i}{K_i/\epsilon
_i+K_t/\epsilon _t}\quad .
\end{equation}
This result suggests that, in addition to the normal components $K={\bf k}%
\bullet {\bf n}$ of the wave vectors, we also introduce ''reduced normal
components'' 
\begin{equation}
\label{mo28}X_i\equiv \frac{K_i}{n_i^2}=\frac{\varepsilon _0}{\varepsilon _i}%
K_i\quad ,\quad X_t\equiv \frac{K_t}{n_t^2}=\frac{\epsilon _0}{\epsilon _t}%
K_t\quad ,
\end{equation}
in which case the coefficients (\ref{mo27}) can be written as 
\begin{equation}
\label{mo29}b_r=\frac{X_i-X_r}{X_i+X_r}\quad ,\quad b_t=\frac{2X_i}{X_i+X_r}%
\quad .
\end{equation}

This notational convention will greatly simplify the computational work in
the proofs of orthogonality of the various modes.

\section{Strategy for proving orthogonality relations}
\label{Strategy}
We will make use of the following formulas: Let $\rho $ be a complex number.
Then 
\begin{equation}
\label{o2}\int\limits_{-\infty }^0dx\;e^{i\rho x}=\frac{-i}\rho \quad ,\quad 
\text{for\ }Im\rho <0\quad ;\quad \int\limits_0^\infty dx\;e^{i\rho x}=\frac
i\rho \quad ,\quad \text{for\ }Im\rho >0\quad .
\end{equation}
As a special case we have 
\begin{equation}
\label{o1}\lim _{\epsilon \rightarrow 0+0}\int\limits_{-\infty
}^0dx\;e^{i\left( k-i\epsilon \right) x}={\cal P}\frac{-i}k+\pi \delta
\left( k\right) \quad ;\quad \lim _{\epsilon \rightarrow
0+0}\int\limits_0^\infty dx\;e^{i\left( k+i\epsilon \right) x}={\cal P}\frac
ik+\pi \delta \left( k\right) \quad .
\end{equation}

As the computational work required to prove orthogonality of triple modes is
considerable, we seek a path of minimal effort. It turns out that a
good strategy goes as follows:

We first prove orthogonality of {\bf electric} fields associated with TE
modes, both for co- and counterpropagating modes; this is done in section 
\ref{TEelectric}. Next we prove orthogonality of {\bf magnetic} fields
associated with TM modes, also for co- and counterpropagating modes; this
will be done in section \ref{TMmagnetic}. Finally, we show that electric as
well as magnetic fields associated with one TE and one TM mode always have a
vanishing orthogonality integral. This will be done in section \ref{TETM}.

After that, in section \ref{SurfContribution} we will prove the equations 
\begin{equation}
\label{str1}\int d^3x\left\{ n^2\left( {\bf x}\right) {\bf E}^{*}\left( {\bf %
x}\right) \bullet {\bf E}^{\prime }\left( {\bf x}\right) -c^2{\bf B}%
^{*}\left( {\bf x}\right) \bullet {\bf B}^{\prime }\left( {\bf x}\right)
\right\} =0\quad , 
\end{equation}
and
\begin{equation}
\label{str1a}
\int d^3x\left\{ n^2\left( x\right) {\bf E}\left( x\right)
\bullet {\bf E}^{\prime }\left( x\right) +c^2{\bf B}\left( x\right) \bullet 
{\bf B}^{\prime }\left( x\right) \right\} =0\quad , 
\end{equation}
for all possible TE and TM modes. With the help of (\ref{str1}) we can then
immediately find the remaining orthogonality relations: Applying (\ref{str1}%
) to the results from section \ref{TEelectric} proves orthogonality of {\bf %
magnetic} fields associated with TE modes; while (\ref{str1}) applied to the
results of section \ref{TMmagnetic} yields orthogonality of {\bf electric}
fields associated with TM modes. Together with (\ref{str1a}) we then 
have completed our task of proving all possible orthogonality relations.

One might wonder why we choose a strategy as contrived as the one above. The
reason is simply that this approach minimizes the computational work to be
done to prove all the relations discussed above; any other strategy would
yield the same results, but would require more effort.

\section{TE modes -- orthogonality of electric fields}
\label{TEelectric}
\subsection{Copropagating TE electric fields}

Consider the electric fields of two TE modes%
\begin{equation}
\label{emodes1a}
{\bf E}\left( {\bf k}_i,1,{\bf x}\right) =\left\{ 
\begin{array}{ccc}
{\bf e}\left[ e^{i{\bf k}_i\bullet {\bf x}}+a_re^{i{\bf k}_r\bullet {\bf x}%
}\right] & , & {\bf x\in }\TeXButton{H}{\mathbb{H}}_i \\ {\bf e}a_te^{i{\bf k%
}_t\bullet {\bf x}} & , & {\bf x\in }\TeXButton{H}{\mathbb{H}}_t 
\end{array}
\quad ,\right. 
\end{equation}
and%
\begin{equation}
\label{emodes1b}
{\bf E}^{\prime }\left( {\bf k}_i^{\prime },1,{\bf x}\right) =\left\{ 
\begin{array}{ccc}
{\bf e}^{\prime }\left[ e^{i{\bf k}_i^{\prime }\bullet {\bf x}}+a_r^{\prime
}e^{i{\bf k}_r^{\prime }\bullet {\bf x}}\right] & , & {\bf x\in }%
\TeXButton{H}{\mathbb{H}}_i \\ {\bf e}^{\prime }a_t^{\prime }e^{i{\bf k}%
_t^{\prime }\bullet {\bf x}} & , & {\bf x\in }\TeXButton{H}{\mathbb{H}}_t 
\end{array}
\right. \quad , 
\end{equation}
where both modes are copropagating, i.e. 
\begin{equation}
\label{or1}sgn\left( K_i\right) =sgn\left( K_i^{\prime }\right) \quad .
\end{equation}
Then
\begin{equation}
\label{or2}\int d^3x\;n^2\left( {\bf x}\right) \,{\bf E}\left( 
\mathbf{k}_i,1,{\bf x}
\right) ^{*}\bullet {\bf E}^{\prime }\left( 
\mathbf{k}'_i,1,{\bf x}\right) =\left(
2\pi \right) ^3n_i^2\;\delta \left( {\bf k}_i-{\bf k}_i^{\prime
}\right) \quad ,
\end{equation}
and the result holds for both combinations ${\bf E}_L, {\bf E}'_L$ and
${\bf E}_R, {\bf E}'_R$ of subscripts.  We see that the square of the
refractive index serves as a weight function in the integrand. On the
right hand side (RHS), $n_i$ is the refractive index in the common
half-space $\mathbb{H}_i$ of the two incident components of the triple
modes.

\underline{Proof:}

We perform the proof for two modes incident from the left. The case of
incident waves coming from the right is very similar. If eqs
(\ref{emodes1a}, \ref{emodes1b}) are inserted into eq (\ref{or2}), the
integration over $x_1$ and $x_2$ can be performed, which results in a
delta function
\begin{equation}
\label{or4a}\left( 2\pi \right) ^2\delta \left( k_1^{\prime }-k_1\right)
\delta \left( k_2^{\prime }-k_2\right) \equiv \left( 2\pi \right) ^2\delta
^{\left( 2\right) }\left( k_{\parallel }^{\prime }-k_{\parallel }\right) 
\end{equation}
 over the parallel components of the wave vectors. 
Thus we can replace ${\bf k}_{\parallel }^{\prime }$ by 
${\bf k}_{\parallel }$ in all
expressions in the sequel. We write this result as 
\begin{equation}
\label{or5}\int d^3x\,n^2\left( {\bf x}\right) \,{\bf E}\left( {\bf x}%
\right) ^{*}\bullet {\bf E}^{\prime }\left( {\bf x}\right) =\left( {\bf e}%
\bullet {\bf e}^{\prime }\right) \left( 2\pi \right) ^2\delta ^{\left(
2\right) }\left( {\bf k}_{\parallel }^{\prime }-{\bf k}_{\parallel }\right)
\times I\quad , 
\end{equation}
where 
\begin{equation}
\label{or6}
\begin{array}{c}
I=\int\limits_{-\infty }^0 dz \; n_i^2
\left[ 
e^{iz(K'_i-K_i)} + a_r^* e^{iz(K'_i+K_i)} +  
a'_r e^{iz(-K'_i-K_i)} + a_r^* a'_r e^{iz(-K'_i + K_i)}
\right] \; + \\ 
+ \; \int\limits_0^\infty dz \; n_t^2 \; a_t^{*}a'_t e^{iz(K'_t-K_t^*)}
\quad ,
\end{array}
\end{equation}
and we have omitted reference to the wave vectors $\mathbf{k}_i$ and
the polarizations $1,2$ in $\mathbf{E}(\mathbf{x})$.
Now we apply theorems (\ref{o2}--\ref{o1}) to equation (\ref{or6}).
Since $K_i$, $K_i^{\prime }$ are always real, the first integral can
be treated using theorem (\ref{o1}), resulting in two contributions
from the principal part and the delta function. The second integral is
more involved: If both $K_t$ and $K_t^{\prime }$ are real, then
theorem (\ref{o1}) gives
\begin{equation}
\label{or7}\int\limits_0^\infty dz\;e^{iz\left( K_t^{\prime }-K_t^{*}\right)
}={\cal P}\frac i{K_t^{\prime }-K_t}+\pi \delta \left( K_t^{\prime
}-K_t\right) \quad . 
\end{equation}
There are three further cases: $K_t^{\prime }$ real, $K_t=i\kappa _t$; $%
K_t^{\prime }=i\kappa _t^{\prime }$, $K_t$ real; $K_t=i\kappa _t$, $%
K_t^{\prime }=i\kappa _t^{\prime }$. Since the incident wave comes from the
left, the imaginary parts $\kappa _t$, $\kappa _t^{\prime }$ must be
positive in each case, and hence the condition that $Im\left( K_t^{\prime
}-K_t^{*}\right) >0$ is always fulfilled, so that we can apply theorem (\ref
{o2}): 
\begin{equation}
\label{or8}\int\limits_0^\infty dz\;e^{iz\left( K_t^{\prime }-K_t^{*}\right)
}={\cal P}\frac i{K_t^{\prime }-K_t^{*}}\quad . 
\end{equation}
In order to treat all four cases simultaneously, we note that the complex 
delta function $\delta \left( K_t^{\prime }-K_t^{*}\right) $ is non-zero
only in the first case discussed above, i.e. when both $K_t^{\prime }$ and $%
K_t$ are real. Hence, results (\ref{or7}) and (\ref{or8}) can be summarized
into one formula: 
\begin{equation}
\label{or9}\int\limits_0^\infty dz\;e^{iz\left( K_t^{\prime }-K_t^{*}\right)
}={\cal P}\frac i{K_t^{\prime }-K_t^{*}}+\pi \delta \left( K_t^{\prime
}-K_t^{*}\right) \quad , 
\end{equation}
where it is understood that the $\delta $-function contributes only
for real argument, and vanishes otherwise. We can now compute
expression (\ref{or6}):
\begin{equation}
\label{or10}
\begin{array}{c}
I=n_i^2\left\{ 
{\cal P}\frac{-i}{K_i^{\prime }-K_i}+a_r^{*}{\cal P}\frac{-i}{K_i^{\prime
}+K_i}+a_r^{\prime }{\cal P}\frac i{K_i^{\prime }+K_i}+a_r^{\prime }a_r^{*}%
{\cal P}\frac i{K_i^{\prime }-K_i}\right\} \quad + \\ +\;n_t^2a_t^{\prime
}a_t^{*} 
{\cal P}\frac i{K_t^{\prime }-K_t^{*}}\quad + \\ +\quad n_i^2\pi \left\{
\delta \left( K_i^{\prime }-K_i\right) +a_r^{*}\delta \left( K_i^{\prime
}+K_i\right) +a_r^{\prime }\delta \left( K_i^{\prime }+K_i\right)
+a_r^{\prime }a_r^{*}\delta \left( K_i^{\prime }-K_i\right) \right\} \quad +
\\ 
+\quad n_t^2\pi a_t^{\prime }a_t^{*}\delta \left( K_t^{\prime
}-K_t^{*}\right) \quad , 
\end{array}
\end{equation}
where we already have split the principal parts from the delta functions.
We now must insert (\ref{reftrans1}) for $a_r$. A long
computation then shows that all principal parts cancel each other: To see
this we use 
\begin{equation}
\label{or13}
\begin{array}{c}
\left( K_t^{\prime }-K_t^{*}\right) \left( K_t^{\prime }+K_t^{*}\right) 
=\left( \omega ^{\prime 2}-\omega ^2\right) \mu _0\epsilon _t\quad  
\end{array}
\end{equation}
to derive 
\begin{equation}
\label{or15}\frac{n_i^2}{\left( K_i^{\prime }-K_i\right) \left( K_i^{\prime
}+K_i\right) }=\frac{n_t^2}{\left(
K_t^{\prime }-K_t^{*}\right) \left( K_t^{\prime }+K_t^{*}\right) }\quad . 
\end{equation}
If the last equation is used in (\ref{or10}),
 all contributions from principal parts cancel out. Since both 
$K_i^{\prime }$ and $K_i$ are positive in the present case, 
the $\delta \left( K_i^{\prime }+K_i\right) 
$-functions in ({\ref{or10}) vanish, and we obtain
\begin{equation}
\label{or19}I=n_i^2\pi \left\{ 1+a_r^{\prime }a_r^{*}\right\} \delta \left(
K_i^{\prime }-K_i\right) +n_t^2\pi a_t^{\prime }a_t^{*}\delta \left(
K_t^{\prime }-K_t^{*}\right) \quad . 
\end{equation}
In this expression, vanishing of the second delta function implies the 
vanishing of the first. Thus, the only occasion at
which $I$ is not zero occurs when both $K_t^{\prime }$ and $K_t$
are real. Then we can express $\delta \left( K_t^{\prime }-K_t\right) $
in terms of $\delta \left( K_i^{\prime }-K_i\right) $,
\begin{equation}
\label{or22}\delta \left( K_t^{\prime }-K_t\right) =\frac{n_i^2}{n_t^2}\frac{%
\left| K_t\right| }{\left| K_i\right| }\delta \left( K_i^{\prime
}-K_i\right) \quad , 
\end{equation}
where we have used the fact that 
$K'_t, K_t$ as well as $K_i, K'_i$ have the same sign.
(\ref{or19}) now becomes 
\begin{equation}
\label{or26}I=\pi n_i^2\left\{ 1+a_r^{\prime }a_r^{*}+a_t^{\prime }a_t^{*}%
\frac{\left| K_t\right| }{\left| K_i\right| }\right\} \delta \left(
K_i^{\prime }-K_i\right) \quad . 
\end{equation}
Since this expression is non-vanishing only for $K_i^{\prime }=K_i$, we can
omit the prime's in the curly brackets. Furthermore $K_t, K_i$ have the 
same sign,
hence the curly brackets yield 
\begin{equation}
\label{or26b}1+a_ra_r^{*}+a_ta_t^{*}\frac{K_t}{K_i}=2\quad . 
\end{equation}
Finally, therefore, 
\begin{equation}
\label{or26c}I=2\pi n_i^2\delta \left( K_i^{\prime }-K_i\right) \quad . 
\end{equation}
If (\ref{or26c}) is combined with (\ref{or5}) we arrive at the result (\ref%
{or2}).

The same result holds for two copropagating modes incident from the
right: Again, the square of the index of refraction $n^2\left( {\bf
x}\right) $ emerges as weight in the integral, while now on the RHS 
$n_i^2$ is equal to $1$.  \hfill $\blacksquare$

\subsection{Counterpropagating TE electric fields}

In a similar way we prove the orthogonality of two counter-propagating TE
modes. It is sufficient to assume that the field $\bf E$ is incoming from 
the left, while $\bf E'$ is incident from the right:

Consider the electric fields of two {\bf counter}-propagating TE modes
\begin{equation}
\mathbf{E}_L\left( {\bf k}_i,1,{\bf x}\right) =\left\{ 
\begin{array}{ccc}
{\bf e}\left[ e^{i{\bf k}_i\bullet {\bf x}}+a_re^{i{\bf k}_r\bullet {\bf x}%
}\right] & , & z<0 \\ {\bf e}a_te^{i{\bf k%
}_t\bullet {\bf x}} & , & z\ge 0 
\end{array}
\quad ,\right. 
\end{equation}
and
\begin{equation}
{\bf E}_R^{\prime }\left( {\bf k}_i^{\prime },1,{\bf x}\right) =\left\{ 
\begin{array}{ccc}
{\bf e}^{\prime }a_t^{\prime }e^{i{\bf k}_t^{\prime }\bullet {\bf x}} & , & 
z<0 \\ {\bf e}^{\prime }\left[ e^{i{\bf k}%
_i^{\prime }\bullet {\bf x}}+a_r^{\prime }e^{i{\bf k}_r^{\prime }\bullet 
{\bf x}}\right] & , & z\ge 0 
\end{array}
\right. \quad ,
\end{equation}
where $K_i>0$ and $K'_i\le 0$. Then 
\begin{equation}
\label{or28}\int d^3x\;n^2\left( {\bf x}\right) \,{\bf E}_L\left( 
\mathbf{k}_i,1,{\bf x}
\right) ^{*}\bullet {\bf E}_R^{\prime }\left( 
\mathbf{k}'_i,1,{\bf x}\right) =0\quad . 
\end{equation}

\underline{Proof:}

Following the same course as in the previous section we find that 
\begin{equation}
\label{or29}
\int d^3x\,n^2\left( {\bf x}\right) \,{\bf E}\left( {\bf x}
\right) ^{*}\bullet {\bf E}^{\prime }\left( {\bf x}\right) =\left( 2\pi
\right) ^2\left( {\bf e}\bullet {\bf e}^{\prime }\right) \delta ^{\left(
2\right) }\left( {\bf k}_{\parallel }-{\bf k}_{\parallel }^{\prime }\right)
\times I\quad , 
\end{equation}
where the quantity $I$ now takes the form 
\begin{equation}
\label{or30}
\begin{array}{c}
I=n_0^2\int\limits_{-\infty }^0dz\left[ a_t^{\prime }e^{iz\left( K_t^{\prime
}-K_i\right) }+a_t^{\prime }a_r^{*}e^{iz\left( K_t^{\prime }+K_i\right)
}\right] + \\ 
+\int\limits_0^\infty dz\left[ a_t^{*}e^{iz\left( K_i^{\prime
}-K_t^{*}\right) }+a_t^{*}a_r^{\prime }e^{iz\left( -K_i^{\prime
}-K_t^{*}\right) }\right] \quad . 
\end{array}
\end{equation}
Since the result is non-vanishing only for ${\bf k}_{\Vert }^{\prime }={\bf k%
}_{\Vert }$, we have ${\bf e}\bullet {\bf e}^{\prime }=1$. 
Integration over $z$ and splitting into principal
part and delta functions using eqs (\ref{o2}--\ref{o1}) gives for the
principal-part-contribution to (\ref{or30}) 
\begin{equation}
\label{or31}
\begin{array}{c}
I_{
\text{princ}}=\frac{4iK_i}{\left( K_i^{\prime }+K_t^{\prime }\right) \left(
K_i+K_t^{*}\right) }\frac 1{K_t^{\prime 2}-K_i^2}\times \\ \times \left\{
-n_1^2\frac{K_i^{\prime }\left( K_t^{\prime }+K_t^{*}\right) }{K_t^{\prime
2}-K_i^2}+n_2^2\frac{K_i^{\prime }\left( K_t^{\prime }+K_t^{*}\right) }{%
K_i^{\prime 2}-K_t^{*2}}\right\} \quad , 
\end{array}
\end{equation}
while the delta contributions to (\ref{or30}) all vanish, since $K_i$ and 
$K'_i$ have now opposing sign. Thus, eq (\ref{or31}) can be rewritten as 
\begin{equation}
\label{or31a}I_{\text{princ}}=\frac{4iK_iK_i^{\prime }\left( K_t^{\prime
}+K_t^{*}\right) }{\left( K_i^{\prime }+K_t^{\prime }\right) \left(
K_i+K_t^{*}\right) }\left\{ -\frac{n_1^2}{K_t^{\prime 2}-K_i^2}+\frac{n_2^2}{%
K_i^{\prime 2}-K_t^{*2}}\right\} \quad . 
\end{equation}
Using the fact that 
\begin{equation}
\label{or32}K_i^{\prime 2}-K_t^{*2}=\frac{\varepsilon _2}{\varepsilon _1}%
\left( K_t^{\prime 2}-K_i^2\right) \quad , 
\end{equation}
and 
\begin{equation}
\label{or33}n_2^2=\frac{\varepsilon _2}{\varepsilon _1}n_1^2\quad , 
\end{equation}
the curly brackets are seen to vanish, which proves the result. 
\hfill $\blacksquare$

\section{TM modes -- orthogonality of magnetic fields}

\label{TMmagnetic}
Here we prove analogous relations for TM modes. Since we start with
magnetic rather than electric fields, we normalize the incoming
magnetic field ${\bf B}_i$ in such a way that the magnitude of the
associated electric field ${\bf E}_i$ is $1$,
\begin{equation}
\label{pre1}\left| {\bf B}_i\right| =\frac{\left| {\bf k}_i\right| }\omega
\left| {\bf E}_i\right| =\frac 1c\quad . 
\end{equation}
Hence, the incoming magnetic plane wave takes the form 
\begin{equation}
\label{pre2}
{\bf B}_i=\frac1{c}\mathbf{e}e^{i{\bf k}_i\bullet {\bf x}}\quad ,\quad
|\mathbf{e}| = 1 \quad ,
\quad  [\mathbf{e}]=[\mathbf{E}] \quad .
\end{equation}

\subsection{Copropagating TM magnetic fields}

Consider the {\bf magnetic} fields of two TM modes
$$
{\bf B}\left( {\bf k}_i,2,{\bf x}\right) =\left\{ 
\begin{array}{ccc}
\frac1c{\bf e}\left[ e^{i{\bf k}_i
\bullet {\bf x}}+b_re^{i{\bf k}_r\bullet {\bf x}
}\right] & , & {\bf x\in }\mathbb{H}_i \\ 
\frac1c{\bf e}b_te^{i{\bf k}_t
\bullet {\bf x}} & , & {\bf x\in }\mathbb{H}_t 
\end{array}
\quad ,\right. 
$$
and
$$
{\bf B}^{\prime }\left( {\bf k}_i^{\prime },2,{\bf x}\right) =\left\{ 
\begin{array}{ccc}
\frac1c{\bf e}^{\prime }\left[ e^{i{\bf k}_i^{\prime }
\bullet {\bf x}}+b_r^{\prime
}e^{i{\bf k}_r^{\prime }\bullet {\bf x}}\right] & , & {\bf x\in }
\mathbb{H}_i \\ 
\frac1c{\bf e}^{\prime }b_t^{\prime }e^{i{\bf k}
_t^{\prime }\bullet {\bf x}} & , & {\bf x\in }\TeXButton{H}{\mathbb{H}}_t 
\end{array}
\right. \quad , 
$$
where both modes are copropagating, i.e. 
\begin{equation}
\label{or34}sgn\left( K_i\right) =sgn\left( K_i^{\prime }\right) \quad . 
\end{equation}
Then
\begin{equation}
\label{or35}\int d^3x\;{\bf B}\left( 
\mathbf{k}_i,2,{\bf x}\right) ^{*}\bullet {\bf B}
^{\prime }\left( 
\mathbf{k}'_i,2,{\bf x}\right) =\frac{{\bf e}^2}{c^2}\left( 2\pi \right)
^3\delta \left( {\bf k}_i-{\bf k}_i^{\prime }\right) \quad ,
\end{equation}
and the result holds for both combinations ${\bf B}_L, {\bf B}_L$ and
${\bf B}_R, {\bf B}_R$ of subscripts. 

{\bf Remark :}\quad It is important to note that, in contrast to the
electric case (\ref{or2}), in the magnetic case {\bf no} refractive
index appears in the integral (and neither on the RHS of the above
equation). The same is true for counterpropagating modes.

\underline{Proof:}

We proceed similar to the proofs given above. In a first step,
we perform the integration over the coordinates ${\bf x}_{\Vert }=\left(
x_1,x_2\right) $ in (\ref{or35}), giving the intermediate result 
\begin{equation}
\label{or35a}\int d^3x\;{\bf B}\left( {\bf x}\right) ^{*}\bullet {\bf B}
^{\prime }\left( {\bf x}\right) = 
\left(\frac{2\pi}c\right)^2\left( {\bf e}\bullet 
{\bf e}^{\prime }\right) \delta ^{\left( 2\right) }\left( {\bf k}_{\Vert }-
{\bf k}_{\Vert }^{\prime }\right) \times I\quad . 
\end{equation}
Due to the delta factor we have ${\bf e}\bullet {\bf e}'=1$.
As before, the expression $I$ can be written as the sum of two
parts $I_{\text{delta}}+I_{\text{princ}}$, involving the delta functions and
the principal values, respectively. The principal part contribution  vanishes
as in the previous computations, $I_{\text{princ}}=0$.
The delta contribution is 
\begin{equation}
\label{or38}I_{\text{delta}}=\pi \left\{ b_t^{*}b_t^{\prime }\delta \left(
K_t^{\prime }-K_t\right) +\delta \left( K_i^{\prime }-K_i\right)
+b_r^{\prime }b_r^{*}\delta \left( K_i^{\prime }-K_i\right) \right\} \quad , 
\end{equation}
where the first term involving $\delta \left( K_t^{\prime }-K_t\right) $ is
present only if both $K_t$ and $K_t^{\prime }$ are real. If either of them
is imaginary then according to the previous discussion the first term
vanishes.

Next we use the fact that the modes are copropagating, 
\begin{equation}
\label{or39}sgn\left( K_i\right) =sgn\left( K_i^{\prime }\right) =sgn\left(
K_t\right) =sgn\left( K_t^{\prime }\right) \quad , 
\end{equation}
to obtain 
\begin{equation}
\label{or41}\delta \left( K_t^{\prime }-K_t\right) =\frac{X_t}{X_i}\delta
\left( K_i^{\prime }-K_i\right) \quad , 
\end{equation}
provided that both $K_t$ and $K_t^{\prime }$ are real. The quantities
$X$ are defined in eq (\ref{mo28}).

Now, three cases must be distinguished: $\left( 1\right) $, both $K_t$ and $%
K_t^{\prime }$ are real; $\left( 2\right) $, both $K_t$ and $K_t^{\prime }$
are imaginary; $\left( 3\right) $, one of them is real and the other one is
imaginary. We first deal with case $\left( 3\right) $: We have the equation 
\begin{equation}
\label{or42}K_t^{\prime 2}-K_t^2=\frac{\epsilon _t}{\epsilon _i}\left(
K_i^{\prime 2}-K_i^2\right) \quad . 
\end{equation}
In case $\left( 3\right) $, the LHS is always nonvanishing, and so is the
RHS. But this means that certainly 
\begin{equation}
\label{or43}K_i\neq K_i^{\prime }\quad , 
\end{equation}
hence all delta functions involving $K_i$ and $K_i^{\prime }$ vanish
in (\ref {or38}). Furthermore, the first delta function involving
$K_t$ and $ K_t^{\prime }$ vanishes naturally in this case, as was
explained in section \ref{TEelectric}. It follows that the
orthogonality integral must vanish in this case, and so eq (\ref{or35})
is confirmed.

Now we turn to case $\left( 1\right) $, where $K_t$ and $K_t^{\prime }$ are
real. Here we find on using (\ref{or41}) that 
\begin{equation}
\label{or44}I_{\text{delta}}=\pi \delta \left( K_i^{\prime }-K_i\right)
\left\{ b_t^2\frac{X_t}{X_i}+1+b_r^2\right\} \quad . 
\end{equation}
Direct computation now shows that the curly brackets yield $2$, and
therefore 
\begin{equation}
\label{or45}I_{\text{delta}}=2\pi \delta \left( K_i^{\prime }-K_i\right)
\quad . 
\end{equation}
(\ref{or45}) together with (\ref{or35a}) now confirms (\ref{or35}).

Finally, we consider the case where both $K_t$ and $K_t^{\prime }$ are
imaginary, and hence are associated with evanescent modes. Now we have
$ K_t^{*}=-K_t$, and (\ref{or38}) gives
\begin{equation}
\label{or46}I_{\text{delta}}=\pi \delta \left( K_i^{\prime }-K_i\right)
\left\{ 1+\left| b_r\right| ^2\right\} \quad . 
\end{equation}
But 
\begin{equation}
\label{or47}\left| b_r\right| ^2=\frac{\left| X_i-X_t\right| ^2}{\left|
X_i+X_t\right| ^2}=1\quad , 
\end{equation}
due to the fact that $X_i$ is real, and $X_t$ is purely imaginary, 
$X_t^{*}=-X_t$. Thus, again we find (%
\ref{or45}) for $I_{\text{delta}}$, which confirms (\ref{or35}). 
\hfill $\blacksquare$

\subsection{Counterpropagating TM magnetic fields}

This case proceeds exactly analogous to the previous results, so we quote
the result without explicit proof:
The {\bf magnetic} fields of two counterpropagating TM modes, 
\begin{equation}
\label{or48}sgn\left( K_i\right) =-sgn\left( K_i^{\prime }\right) \quad , 
\end{equation}
are orthogonal, 
\begin{equation}
\label{or49}\int d^3x\;{\bf B}_L\left( 
\mathbf{k}_i,2,{\bf x}\right) ^{*}\bullet {\bf B}_R
^{\prime }\left( 
\mathbf{k}'_i,2,{\bf x}\right) =0\quad . 
\end{equation}

\section{One TE- , one TM mode}
\label{TETM}
Finally, we show that a pair of modes such that one member is TE and the
other one is TM, is always orthogonal. This holds for both electric and
magnetic fields:

\begin{eqnarray}
\int d^3x\;n^2(x) \; {\bf E}\left( 
\mathbf{k}_i,1,{\bf x}\right)^{*}
\bullet {\bf E}'\left( 
\mathbf{k}'_i,2,{\bf x}\right)  & = &  0  \label{or50}\\
\int d^3x\;{\bf B}\left( 
\mathbf{k}_i,1,\mathbf{x}\right) ^{*}
\bullet {\bf B}'\left( 
\mathbf{k}'_i,2,{\bf x}\right)  & = &  0   \label{or50aa} \quad .
\end{eqnarray}
These results hold for all combinations of subscripts $LL$, $LR$, $RL$
and $RR$.

\underline{Proof:}

From the results above, we know that each of the integrals $J$ in eqs 
(\ref{or50}, \ref{or50aa}) has the form
\begin{equation}
\label{or51}J=\left( 2\pi \right) ^2\delta ^{\left( 2\right) }\left( {\bf k}%
_{\Vert }-{\bf k}_{\Vert }^{\prime }\right) \times I\quad , 
\end{equation}
which follows from performing the integration in (\ref{or50}) over the two
coordinates ${\bf x}_{\Vert }$. Because of the delta factor, both wave
vectors lie in the same plane of incidence. First, consider the electric
case in (\ref{or50}), and assume that ${\bf E}$ is TE, and ${\bf E}^{\prime
} $ is TM. Then ${\bf E}$ is perpendicular to the plane of incidence, while $%
{\bf E}^{\prime }$ lies in the plane, hence the LHS of (\ref{or50})
vanishes. -- The same argument immediately applies to the magnetic case. 
\hfill $\blacksquare$

\section{A theorem linking electric and magnetic orthogonality}
\label{SurfContribution}
Now we turn to prove the following equations, 
\begin{equation}
\label{or52a}\int d^3x\left\{ n^2\left( x\right) {\bf E}^{*}\left( x\right)
\bullet {\bf E}^{\prime }\left( x\right) -c^2{\bf B}^{*}\left( x\right)
\bullet {\bf B}^{\prime }\left( x\right) \right\} =0\quad , 
\end{equation}
\begin{equation}
\label{or52b}\int d^3x\left\{ n^2\left( x\right) {\bf E}\left( x\right)
\bullet {\bf E}^{\prime }\left( x\right) +c^2{\bf B}\left( x\right) \bullet 
{\bf B}^{\prime }\left( x\right) \right\} =0\quad , 
\end{equation}
valid for all combinations $LL$, $LR$, $RL$ and $RR$ of subscripts as
well as for both TE and TM polarizations. These equations will
complete our discussion of orthogonality of triple modes.

\textbf{Remark:}\quad Note that the first of these equations involves
one complex conjugate of an electric and a magnetic field,
respectively, while the second equation contains no complex conjugate
fields.

\underline{Proof:}

We first prove (\ref{or52a}).

We start by introducing the explicit time dependence of the modes, 
\begin{equation}
\label{or53}
\begin{array}{c}
{\bf E}\equiv {\bf E}\left( x\right) e^{-i\omega t}\quad ,\quad {\bf B}%
\equiv {\bf B}\left( x\right) e^{-i\omega t}\quad , \\ {\bf E}^{\prime
}\equiv {\bf E}^{\prime }\left( x\right) e^{-i\omega ^{\prime }t}\quad
,\quad {\bf B}^{\prime }\equiv {\bf B}^{\prime }\left( x\right) e^{-i\omega
^{\prime }t}\quad . 
\end{array}
\end{equation}
These time-dependent fields must obey Maxwell's equations, from which we
derive
\begin{equation}
\label{or57}\nabla \bullet \left[ {\bf E}^{*}\times {\bf B}^{\prime }\right]
=i\left\{ \omega ^{\prime }\frac{n^2}{c^2}{\bf E}^{*}\bullet {\bf E}^{\prime
}-\omega {\bf B}^{*}\bullet {\bf B}^{\prime }\right\} \quad . 
\end{equation}
This equation contains a time dependence $\exp i\left( \omega -\omega
^{\prime }\right) t$ on both sides. Integrating over $t$ and 
$\omega'$  yields
\begin{equation}
\label{or58}\frac 1\omega \nabla \bullet \left[ {\bf E}^{*}\left( x\right)
\times {\bf B}^{\prime }\left( x\right) \right] =i\left\{ \frac{n^2}{c^2}%
{\bf E}^{*}\left( x\right) \bullet {\bf E}^{\prime }\left( x\right) -{\bf B}%
^{*}\left( x\right) \bullet {\bf B}^{\prime }\left( x\right) \right\} \quad
, 
\end{equation}
where ${\bf E}\left( x\right) $ etc are again triple mode functions. In order
to prove (\ref{or52a}) we therefore must show that 
\begin{equation}
\label{or59}\int d^3x\;\nabla \bullet \left[ {\bf E}^{*}\left( x\right)
\times {\bf B}^{\prime }\left( x\right) \right] =0\quad , 
\end{equation}
for any two triple modes ${\bf E}\left( x\right) $ and ${\bf B}^{\prime
}\left( x\right) $. To see this we first note that the spatial dependence of
the integrand can be written as 
\begin{equation}
\label{or60a}{\bf E}^{*}\left( x\right) \times {\bf B}^{\prime }\left(
x\right) ={\bf C}\left( z\right) e^{i{\bf x}_{\Vert }\bullet \left( {\bf k}%
_{\Vert }^{\prime }-{\bf k}_{\Vert }\right) }\quad , 
\end{equation}
where ${\bf C}\left( z\right) $ is given by the expression
\begin{equation}
\label{or62}{\bf C}\left( z\right) =e^{-i{\bf x}_{\Vert }\bullet \left( {\bf %
k}_{\Vert }^{\prime }-{\bf k}_{\Vert }\right) }\left[ {\bf E}^{*}\left(
x\right) \times {\bf B}^{\prime }\left( x\right) \right] \quad . 
\end{equation}
A brief investigation shows that $\mathbf{C}(z)$ indeed depends only
on $z$, as suggested by the notation.  Insertion into (\ref{or59})
yields
\begin{equation}
\label{or60}
\int d^3x\;\nabla \bullet \left[ 
{\bf E}^{*}\left( x\right) \times {\bf B}^{\prime }\left( x\right) \right]
=\left( 2\pi \right) ^2\delta ^{\left( 2\right) }\left( {\bf k}_{\Vert
}^{\prime }-{\bf k}_{\Vert }\right) \int dz\left[ \partial _z{\bf C}\left(
z\right) \right] \quad . 
\end{equation}
It follows that we must show the vanishing of the surface integral 
\begin{equation}
\label{or61}\int dz\left[ \partial _z{\bf C}\left( z\right) \right] =0\quad
,\quad \text{for\quad }{\bf k}_{\Vert }^{\prime }={\bf k}_{\Vert }\quad . 
\end{equation}
We see immediately from eq (\ref{or62}) that, if ${\bf E}$ is TE and
${\bf B}^{\prime }$ is TM, then ${\bf C}$ and hence the integral
(\ref{or60}) vanish as required. To show that the same holds if ${\bf
E}$ is TM and ${\bf B}^{\prime }$ is TE, we take the complex conjugate
of eq (\ref{or58}),
\begin{equation}
\label{or63}\frac 1\omega \nabla \bullet \left[ {\bf E}^{*}\left( x\right)
\times {\bf B}^{\prime }\left( x\right) \right] =\left\{ -\frac 1{\omega
^{\prime }}\nabla \bullet \left[ {\bf E}^{\prime *}\left( x\right) \times 
{\bf B}\left( x\right) \right] \right\} ^{*}\quad . 
\end{equation}
Since by assumption, ${\bf E}^{\prime }$ and ${\bf B}$ are parallel, the RHS
of (\ref{or63}) vanishes, which proves the above statement. 

Thus, we are
left to prove the following four cases:

\begin{enumerate}
\item  ${\bf E}$ and ${\bf B}^{\prime }$ TE, copropagating;

\item  ${\bf E}$ and ${\bf B}^{\prime }$ TE, counter-propagating;

\item  ${\bf E}$ and ${\bf B}^{\prime }$ TM, copropagating;

\item  ${\bf E}$ and ${\bf B}^{\prime }$ TM, counter-propagating.
\end{enumerate}

Due to the $\delta ^{\left( 2\right) }\left( {\bf k}_{\Vert }^{\prime }-{\bf %
k}_{\Vert }\right) $ factor in (\ref{or60}), we can assume that both wave
vectors lie in the same plane of incidence. This implies, in particular,
that in cases $\left( 1\right) $ and $\left( 2\right) $, the magnetic field $%
{\bf B}^{\prime }$ lies in the common plane of incidence, while ${\bf E%
}$ is perpendicular to it; and in cases $\left( 3\right) $ and $\left(
4\right) $, the electric field ${\bf E}$ lies in the plane of incidence,
while ${\bf B}^{\prime }$ is perpendicular.

Let us now treat case $\left( 1\right) $: For the sake of simplicity, we
assume that both modes propagate from the left to the right, i.e. $%
K_i,K_i^{\prime }>0$. We decompose the divergence in eq (\ref{or60}) into 
 parallel and normal derivatives,
\begin{equation}
\label{or68}\nabla \bullet \left( {\bf E}^{*}\times {\bf B}%
^{\prime }\right) =\nabla _{\Vert }\bullet \left( {\bf E}%
^{*}\times {\bf B}^{\prime }\right) _{\Vert }+\partial
_3\left( {\bf E}^{*}\times {\bf B}^{\prime }\right) _3\quad . 
\end{equation}
A computation shows that integration of the first term on the RHS of 
(\ref{or68}) over the spatial coordinates vanishes because of property
(\ref{or63}). In the second term, all terms involving delta functions
vanish. Evaluating the principal value contributions gives
\begin{equation}
\label{or73}
\begin{array}{c}
\int d^3x\;\omega ^{\prime }\partial _3\left( 
{\bf E}^{*}\times {\bf B}^{\prime }\right) _3=\left( 2\pi \right) ^2\delta
^{\left( 2\right) }\left( {\bf k}_{\Vert }^{\prime }-{\bf k}_{\Vert }\right)
\omega ^{\prime }\cdot \left\{ K_i^{\prime }\left\{ 1-a_r^{\prime
}+a_r^{*}-a_r^{\prime }a_r^{*}\right\} -\right. \\ \left. -K_t^{\prime
}a_t^{\prime }a_t^{*}\right\} \quad . 
\end{array}
\end{equation}
On account of
\begin{equation}
\label{or74}1-a_r^{\prime }+a_r^{*}-a_r^{\prime }a_r^{*}=\frac{K_t^{\prime }%
}{K_i^{\prime }}a_t^{\prime }a_t^{*}\quad , 
\end{equation}
we find that the RHS of (\ref{or73}) must vanish. The
statement that integral (\ref{or60}) vanishes is therefore confirmed, for
case $\left( 1\right) $.

Now we briefly discuss the remaining three cases: Case $(2)$ proceeds 
along the same lines as case $(1)$. 
To prove cases $(3)$ and $(4)$ we start with 
\begin{equation}
\label{or81}\nabla \bullet \left( {\bf E}^{*}\times {\bf B}^{\prime }\right)
=-\nabla \bullet \left( {\bf B}^{\prime }\times {\bf E}^{*}\right) \quad . 
\end{equation}
On the RHS, we express ${\bf E}$ by the associated magnetic field ${\bf B}$,
which is TM, 
\begin{equation}
\label{or82}{\bf E}_i=-\frac{{\bf k}_i\times {\bf B}_i}{\omega \mu
_0\epsilon _i}\quad ,\quad {\bf E}_r=-\frac{{\bf k}_r\times {\bf B}_r}{%
\omega \mu _0\epsilon _i}\quad ,\quad {\bf E}_t=-\frac{{\bf k}_t\times {\bf B%
}_t}{\omega \mu _0\epsilon _t}\quad . 
\end{equation}
Subsequent evaluation of the integral then yields a factor $\left( 2\pi
\right) ^2\delta ^{\left( 2\right) }\left( {\bf k}_{\Vert }^{\prime }-{\bf k}
_{\Vert }\right) $ times an expression $\frac 1{\mu _0}f\left(
X_i,X_t,X_i^{\prime },X_t^{\prime },b_r,b_t,b_r^{\prime },b_t^{\prime
}\right) $, where $f$ as a function of the arguments $X$ and $b$ 
has the same functional form as the analogous 
expression in the TE case as a function of respective TE quantities $K$ and 
$a$,
\begin{equation}
\label{or83}f\left( X_i,X_t,X_i^{\prime },X_t^{\prime },b_r,b_t,b_r^{\prime
},b_t^{\prime }\right) \leftrightarrow f\left( K_i,K_t,K_i^{\prime
},K_t^{\prime },a_r,a_t,a_r^{\prime },a_t^{\prime }\right) \quad . 
\end{equation}
Now, the functional form of the quantities $b$ in terms of $X$ is the
same as the functional form of $a$ in terms of $K$, as can be seen by
comparing eqs (\ref{reftrans1}) and (\ref{mo28}). It follows that we
may transfer all conclusions derived previously for the TE case to the
TM case by making the replacements indicated in the last formula
(\ref{or83}). In particular, vanishing $f$ on the RHS of (\ref{or83})
implies vanishing $f$ on the LHS. In this way, we have traced back
cases $\left(3,4\right) $ to cases $\left(1,2\right) $, which finishes
the proof of (\ref{or52a}).

--- Now we can turn to the proof of eq (\ref{or52b}). 
This proceeds analogous to 
the previous equation (\ref{or52a}), hence we give only an outline:
By using the explicit time dependence (\ref
{or53}) of the modes we can derive the relation 
\begin{equation}
\label{or85}\frac 1\omega \nabla \bullet \left[ {\bf E}\left( x\right)
\times {\bf B}^{\prime }\left( x\right) \right] =i\left\{ \frac{n^2}{c^2}%
{\bf E}\left( x\right) \bullet {\bf E}^{\prime }\left( x\right) +{\bf B}%
\left( x\right) \bullet {\bf B}^{\prime }\left( x\right) \right\} \quad 
\end{equation}
between triple mode functions, which is analogous to eq (\ref{or58}). 
To prove (\ref{or52b}) we therefore must show that 
\begin{equation}
\label{or86}\int d^3x\;\nabla \bullet \left[ {\bf E}\left( x\right) \times 
{\bf B}^{\prime }\left( x\right) \right] =0\quad , 
\end{equation}
which is analogous to (\ref{or59}). The left hand side of the last equation 
can again be cast into a form similar to (\ref{or60}),
\begin{equation}
\label{or87}\int d^3x\;\nabla \bullet \left[ {\bf E}^{*}\left( x\right)
\times {\bf B}^{\prime }\left( x\right) \right] =\left( 2\pi \right)
^2\delta ^{\left( 2\right) }\left( {\bf k}_{\Vert }^{\prime }+{\bf k}_{\Vert
}\right) \int dz\left[ \partial _z{\bf C}\left( z\right) \right] \quad , 
\end{equation}
except that the sign of the second wave vector in the argument of the
delta function now differs.  It follows that we must show the
vanishing of the surface integral
\begin{equation}
\label{or89}\int dz\left[ \partial _z{\bf C}\left( z\right) \right] =0\quad
,\quad \text{for\quad }{\bf k}_{\Vert }^{\prime }=-{\bf k}_{\Vert }\quad , 
\end{equation}
which is analogous to (\ref{or61}). The remainder of the proof proceeds 
along the same lines as the proof for (\ref{or52a}).
This finishes our proof of eqs (\ref{or52a}, \ref{or52b}).
{\hfill $\blacksquare$}

--- The remaining orthogonality relations now follow from
(\ref{or52a}): Suppose that ${\bf E}$ and ${\bf E}^{\prime }$ are
electric fields associated with TM modes; then orthogonality follows
from orthogonality of the magnetic fields as proven in section
\ref{TMmagnetic} and (\ref{or52a}). On the other hand, suppose that
${\bf B}$ and ${\bf B}^{\prime }$ are magnetic fields associated with
TE modes; then their orthogonality follows from orthogonality of the
associated electric fields as proven in section \ref {TEelectric} and
again eq (\ref{or52a}). This finishes our proof of orthogonality
relations.

\section{Normalized modes}
\label{NormalizedModes}
In the developments above, we have normalized the electric field of TE
modes upon unit amplitude of the incoming plane wave, and the magnetic
field of TM modes upon a value numerically equal to $1/c$ for the
incoming plane wave. This choice was motivated by simplification of
the subsequent computational work. Now we finally introduce a
normalization which is best suited for the purpose of diagonalization
of the free field Hamiltonian, as performed below. The new
normalization will affect only TE modes; in this case, eq (\ref{or2})
motivates that we divide these modes by the refractive index $n_i$ of
the medium in the incoming half-space $\mathbb{H}_i$. If the TE mode
under consideration is incoming from the right, the associated
refractive index is actually $1$, so that these modes are
unaffected. Thus, the new normalization effectively changes only the
TE modes incoming from the left.

The new TE modes are thus defined as
\begin{equation}
\label{su8a}
{\bf E}_L\left({\bf k}_i , 1 , {\bf x}\right) \equiv \left\{ 
\begin{array}{ccc}
\frac 1{n_0} {\bf e}\left[ e^{i{\bf k}_i\bullet {\bf x}
}+a_r\cdot e^{i{\bf k}_r\bullet {\bf x}}\right] & , & z<0 \\ 
\frac 1{n_0}{\bf e}\cdot a_t\cdot e^{i{\bf k}_t\bullet 
{\bf x}} & , & z\ge 0 
\end{array}
\right. \quad , 
\end{equation}
and
\begin{equation}
\label{su8b}
{\bf E}_R\left({\bf k}_i , 1 , {\bf x}\right) \equiv \left\{ 
\begin{array}{ccc}
{\bf e}\left[ e^{i{\bf k}_i\bullet {\bf x}
}+a_r\cdot e^{i{\bf k}_r\bullet {\bf x}}\right] & , & z\ge0 \\ 
{\bf e}\cdot a_t\cdot e^{i{\bf k}_t\bullet 
{\bf x}} & , & z< 0 
\end{array}
\right. \quad , 
\end{equation}
with 
\begin{equation}
\label{su10}a_r=\frac{K_i-K_t}{K_i+K_t}\quad ,\quad a_t=\frac{2K_i}{K_i+K_t}
\quad . 
\end{equation}
Furthermore, ${\bf B}\left( {\bf k}_i,1,{\bf x}\right) $ is defined to
be the magnetic field associated with (\ref{su8a}, \ref{su8b}), where
each of the plane wave components of ${\bf B}$ is given in terms of
the plane wave components of $ {\bf E}$ according to
\begin{equation}
\label{su11}{\bf B}_i{\bf =}\frac{{\bf k}_i\times {\bf E}_i}\omega \quad
,\quad {\bf B}_r{\bf =}\frac{{\bf k}_r\times {\bf E}_r}\omega \quad ,\quad 
{\bf B}_t{\bf =}\frac{{\bf k}_t\times {\bf E}_t}\omega \quad . 
\end{equation}
Formulas (\ref{su10}, \ref{su11}) are valid for both left- and
right-incoming modes. This defines the \underline{normalized TE
modes}.

The \underline{normalized TM modes} are given by 
\begin{equation}
\label{su12}
{\bf B}_L\left( {\bf k}_i,2,{\bf x}\right) =\left\{ 
\begin{array}{ccc}
\frac 1c{\bf e}\left[ e^{i{\bf k}_i\bullet {\bf x}}+b_{Lr}\cdot
e^{i{\bf k}_r\bullet
{\bf x}}\right] & , & z< 0 \\
\frac 1c{\bf 
e}\cdot b_{Lt}\cdot e^{i{\bf k}_t\bullet {\bf x}} & , & z\ge 0
\end{array}
\right. \quad ,
\end{equation}
where the reflection and transmission coefficients follow from eqs 
(\ref{mo27}--\ref{mo29}),
\begin{equation}
\label{su13}
b_{Lr}=\frac{K_i-n_0^2 K_t}{K_i+n_0^2 K_t} \quad,\quad 
b_{Lt}=\frac{2K_i}{K_i+n_0^2 K_t} \quad , 
\end{equation}
and for modes incoming from the right,
\begin{equation}
\label{su12aa}
{\bf B}_R\left( {\bf k}_i,2,{\bf x}\right) =\left\{ 
\begin{array}{ccc}
\frac 1c{\bf e}\left[ e^{i{\bf k}_i\bullet {\bf x}}+b_{Rr}\cdot e^{i{\bf k}_r\bullet 
{\bf x}}\right] & , & z\ge 0 \\
\frac 1c{\bf 
e}\cdot b_{Rt}\cdot e^{i{\bf k}_t\bullet {\bf x}} & , & z< 0
\end{array}
\right. 
\end{equation}
with
\begin{equation}
\label{su13aa}
b_{Rr}=\frac{n_0^2 K_i-K_t}{n_0^2 K_i+K_t} \quad,\quad 
b_{Rt}=\frac{2K_i}{K_i+\frac{K_t}{n_0^2}} \quad . 
\end{equation}
The plane wave components of the associated TM electric fields are
given in terms of plane wave components of the TM magnetic fields: For
waves incoming from the left we have
\begin{equation}
\label{su14}
{\bf E}_{Li}=-\frac{c^2}{n_0^2}\frac{{\bf k}_i\times {\bf B}_i}{
\omega }\quad ,\quad 
{\bf E}_{Lr}=-\frac{c^2}{n_0^2}\frac{{\bf k}_r\times {\bf B}_r}{
\omega }\quad ,\quad 
{\bf E}_{Lt}=-c^2\frac{{\bf k}_t\times {\bf B}_t}{\omega }\quad ,
\end{equation}
while for waves incoming from the right,
\begin{equation}
\label{su14a}
{\bf E}_{Ri}=-c^2\frac{{\bf k}_i\times {\bf B}_i}{
\omega }\quad ,\quad 
{\bf E}_{Rr}=-c^2\frac{{\bf k}_r\times {\bf B}_r}{
\omega }\quad ,\quad 
{\bf E}_{Rt}=-\frac{c^2}{n_0^2}\frac{{\bf k}_t\times {\bf B}_t}{\omega
}\quad .
\end{equation}

All orthogonality relations discussed before can now summarized:
\begin{equation}
\label{su15}\int d^3x\;n^2\left( x\right) \;{\bf E}^{*}\left( {\bf k},s,{\bf 
x}\right) \bullet {\bf E}^{\prime }\left( {\bf k}^{\prime },s^{\prime },{\bf 
x}\right) =\left( 2\pi \right) ^3\delta _{ss^{\prime }}\delta \left( {\bf k}
^{\prime }-{\bf k}\right) \quad , 
\end{equation}
\begin{equation}
\label{su16}\int d^3x\;{\bf B}^{*}\left( {\bf k},s,{\bf x}\right) \bullet 
{\bf B}^{\prime }\left( {\bf k}^{\prime },s^{\prime },{\bf x}\right) =\left(
2\pi \right) ^3\frac 1{c^2}\delta _{ss^{\prime }}\delta \left( {\bf k}_i-%
{\bf k}_i^{\prime }\right) \quad , 
\end{equation}
for both combinations $LL$ and $RR$, 
and
\begin{equation}
\label{su15a}
\int d^3x\;n^2\left( x\right) \;{\bf E}_L^{*}\left( {\bf k},s,{\bf
x}\right) \bullet {\bf E}_R^{\prime }\left( {\bf k}^{\prime
},s^{\prime },{\bf x}\right) = 0 \quad ,
\end{equation}
\begin{equation}
\label{su16a}\int d^3x\;{\bf B}_L^{*}\left( {\bf k},s,{\bf x}\right) \bullet 
{\bf B}_R^{\prime }\left( {\bf k}^{\prime },s^{\prime },{\bf x}\right) =
0 \quad .
\end{equation}
Furthermore, eqs (\ref{or52a}, \ref{or52b}) continue to hold.

\section{Expansion of arbitrary {\bf source-free} fields in terms of triple
modes}

The set of triple modes is complete with respect to the source-free
fields, as has been shown in \cite{BB1972}. This means that every
source-free electric / magnetic field can be uniquely expressed as a
linear combination of triple modes.  For an arbitrary free {\bf real} 
electric field ${\bf E}\left( {\bf x},t\right)$ we can make the ansatz
\begin{gather}
\label{su17}
{\bf E}\left( {\bf x,}t\right)  =  \\ =\quad
i\int\limits_{K<0}d^3k\sum_{s=1}^2{\cal E}
(k) \left[u_{Ls}\left( {\bf k}\right) {\bf E}_{L}\left( {\bf k},s,{\bf
x}\right) e^{-i\omega t}-u_{Ls}^{*}\left( {\bf k}\right) {\bf
E}_{L}^{*}\left( {\bf k},s,{\bf x}\right) e^{i\omega t}\right] \nonumber \\
+ \quad
i\int\limits_{K\ge 0}d^3k\sum_{s=1}^2{\cal E}(k)
\left[u_{Rs}\left( {\bf k}\right) {\bf E}_{R}\left( {\bf k},s,{\bf
x}\right) e^{-i\omega t}-u_{Rs}^{*}\left( {\bf k}\right) {\bf
E}_{R}^{*}\left( {\bf k},s,{\bf x}\right) e^{i\omega t}\right] 
\nonumber \quad,
\end{gather}
where 
\begin{equation}
\label{su18}{\cal E}\left( k\right) \equiv \sqrt{\frac{\hbar c\left| {\bf k}%
\right| }{2\epsilon _0\left( 2\pi \right) ^3}}\quad . 
\end{equation}
The associated magnetic field is determined by Maxwell's equation
$\nabla \times {\bf E}=-{\bf \dot B}$, and has the expansion
\begin{gather}
\label{su19}
{\bf B}\left( {\bf x,}t\right)  =  \\ =\quad
i\int\limits_{K<0}d^3k\sum_{s=1}^2{\cal E}
(k) \left[u_{Ls}\left( {\bf k}\right) {\bf B}_{L}\left( {\bf k},s,{\bf
x}\right) e^{-i\omega t}-u_{Ls}^{*}\left( {\bf k}\right) {\bf
B}_{L}^{*}\left( {\bf k},s,{\bf x}\right) e^{i\omega t}\right] \nonumber \\
+ \quad
i\int\limits_{K\ge 0}d^3k\sum_{s=1}^2{\cal E}(k)
\left[u_{Rs}\left( {\bf k}\right) {\bf B}_{R}\left( {\bf k},s,{\bf
x}\right) e^{-i\omega t}-u_{Rs}^{*}\left( {\bf k}\right) {\bf
B}_{R}^{*}\left( {\bf k},s,{\bf x}\right) e^{i\omega t}\right] 
\nonumber \quad.
\end{gather}
We note that the amplitude factor $\mathcal{E}(k)$ is the same as in
formula (\ref{su17}).

\section{Expansion of the Hamiltonian in terms of triple modes}

The Hamiltonian of the source-free radiation field is based on the
classical electromagnetic field energy
\begin{equation}
\label{su23}H=\frac{\epsilon _0}2\int d^3x\;\left[ n^2\left( {\bf x}\right) 
{\bf E}^2\left( {\bf x,}t\right) +c^2{\bf B}^2\left( {\bf x},t\right)
\right] \quad , 
\end{equation}
with {\bf real} fields ${\bf E}$ and ${\bf B}$. If we insert
expansions (\ref {su17}, \ref{su19}) into (\ref{su23}), we obtain the 
\underline{mode expansion of the Hamiltonian}
\begin{align}
H & = \quad \int\limits_{K<0}d^3k
\sum_{s=1}^2\frac{\hbar c\left| {\bf k}\right| }2
\left[u_{Ls}\left({\bf k}\right)u_{Ls}^{*}\left( {\bf k}\right)
+u_{Ls}^{*}\left( {\bf k}\right) u_{Ls}\left( {\bf k}\right)
\right] \nonumber\\
& + \quad \int\limits_{K\ge 0}d^3k
\sum_{s=1}^2\frac{\hbar c\left| {\bf k}\right| } 2\left[u_{Rs}\left(
{\bf k}\right) u_{Rs}^{*}\left( {\bf k}\right) +u_{Rs}^{*}\left( {\bf
k}\right) u_{Rs}\left( {\bf k}\right) \right] \quad . \label{su24}
\end{align}
To arrive at this expression, we had to employ all orthogonality
relations (\ref{su15}--\ref{su16a}) and (\ref{or52a}, \ref{or52b}).

In our current discussion, the coefficients $u_{LR,s}$ are
$c$-numbers, as we are still in the classical domain. In deriving the
mode expansion (\ref{su24}), we have been careful, however, to retain
the order in which terms appear, so that (\ref{su24}) can be
immediately transferred into the quantum regime, by interpreting
$u_{Ls}\left( {\bf k}\right) $, $u_{Rs}^{\dagger }\left( {\bf k}
\right)$ etc. as annihilation and creation operators of triple modes.

\section{Summary}

We have worked out the orthogonality relations for the set of
Carniglia-Mandel triple modes, which provide a set of normal modes for
the source-free electromagnetic field in a background consisting of a
dielectric half-space and the vacuum, respectively. The inherent
computational complexity in this kind of problem creates a demand for
efficient strategies to accomplish this task. In this paper we have
provided this strategy, and furthermore, we have presented a
comprehensive discussion of the proofs for the various distinct cases
in the orthogonality integral in order to serve as a compilation of
technical details for related problems in the field of optics with
possibly even greater technical complexity.

\section{Acknowledgement}

Hanno Hammer wishes to acknowledge support from the Weizmann Institute
of Sciences.


\begin{thebibliography}{99}

\bibitem{Purcell} Purcell, E. M., 1946, {\it Phys.\ Rev.}, {\bf 69},
681.

\bibitem{JaynesCummings1963} Jaynes, E. T., and Cummings, F. W., 1963,
{\it IEEE J.\ Quant.\ Elec.}, {\bf 51}, 89.

\bibitem{Stehle1970} Stehle, P., 1970, {\it Phys.\ Rev.\ A}, {\bf 2},
102.

\bibitem{Barton1970} Barton, G., 1970, {\it Proc.\ Royal Soc.\ London
A}, {\bf 320}, 251.

\bibitem{Hinds1} Hinds, E. A., 1991, Cavity Quantum
Electrodynamics. {\it Advances in Atomic, Molecular and Optical
Physics}, 28, 237--286.

\bibitem{Berman1994} Berman, P. R. (ed.), 1994, {\it Cavity Quantum
Electrodynamics } (San Diego: Academic Press).

\bibitem{CarnigliaMandel1} Carniglia, C. K., and Mandel, L., 1971,
{\it Phys.\ Rev.\ D}, {\bf 3}, 280.

\bibitem{CarnigliaMandel2} Carniglia, C. K., Mandel, L., and Drexhage,
K. H., 1972, {\it J.\ Opt.\ Soc.\ Am.}, {\bf 62}, 479.

\bibitem{JanZak1994} M. Janowicz and W. \.Zakowicz, 1994, {\it
Phys. Rev. A}, {\bf 50}, 4350.

\bibitem{InoueHori2001} Inoue, T., and Hori, H., 2001, {\it
Phys.\ Rev.\ A}, {\bf 63}, 063805.

\bibitem{UrbachRikken1} Urbach, H. P., and Rikken, G. L. J. A., 1998, 
{\it Phys.\ Rev.\ A}, {\bf 57}, 3913.

\bibitem{ZakowiczBledowski1995} \.Zakowicz, W., and Bledowski, A.,
1995, {\it Phys.\ Rev.\ A}, {\bf 52}, 1640.

\bibitem{GlauberLewenstein1991} Glauber, R. J., and Lewenstein, M.,
1991, {\it Phys.\ Rev.\ A}, {\bf 43}, 467.

\bibitem{BB1972} Bialynicki-Birula, I., and Brojan, J. B., 1972, {\it
Phys.\ Rev.\ D}, {\bf 5}, 485.

\bibitem{Jackson} Jackson, J. D., 1999, {\it Classical
Electrodynamics} (New York: John Wiley), 3rd ed.

\bibitem{Wolf} Born, M., and Wolf, E., 1970, {\it Principles of
Optics} (Oxford: Pergamon), 4th ed.

\end{thebibliography}
\end{document}